\documentclass[twocolumn,showpacs,preprintnumbers,amsmath,amssymb]{revtex4}

\usepackage{graphicx}
\usepackage{dcolumn}
\usepackage{bm}

\begin{document}

\preprint{APS/123-QED}

\title{$^{19}$F NMR Investigation of Iron-pnictide Superconductor LaFeAsO$_{0.89}$F$_{0.11}$ }

\author{K. Ahilan$^{1}$, F. L. Ning$^{1}$, T. Imai$^{1,2}$, A. S. Sefat$^{3}$, R. Jin$^{3}$, M.A. McGuire$^{3}$, B.C. Sales$^{3}$, and D. Mandrus$^{3}$}

\affiliation{$^{1}$Department of Physics and Astronomy, McMaster University, Hamilton, Ontario L8S4M1, Canada}
\affiliation{$^{2}$Canadian Institute for Advanced Research, Toronto, Ontario M5G1Z8, Canada}
\affiliation{$^{3}$Materials Science and Technology Division, Oak Ridge National Laboratory, TN 37831, USA}

\date{\today}

\begin{abstract}
We report $^{19}$F NMR investigation of the high temperature superconductor LaFeAsO$_{0.89}$F$_{0.11}$ ($T_{c}\sim 28$~K).  We demonstrate that low frequency spin fluctuations exhibit pseudo gap behavior above $T_c$.  We also deduce the London penetration depth $\lambda$ from NMR line broadening below $T_c$. 
\end{abstract}


\maketitle


The recent discovery of an iron-pnictide superconductor LaFeAsO$_{1-x}$F$_{x}$ ($x=0.05 \sim 0.12$) with a high transition temperature $T_{c}\sim 26$~K by Kamihara et al.  \cite{Kamihara} has generated tremendous interest in the condensed matter and materials physics communities.  When La$^{3+}$ ions are replaced by Nd$^{3+}$ ions, $T_{c}$ exceeds 50~K \cite{RenNd}.  The key building blocks of the RFeAsO$_{1-x}$F$_{x}$ system (R = La, Nd etc.) are two dimensional sheets of Fe square lattice, as shown in Fig.1.  RO$_{1-x}$F$_{x}$ charge reservoir layers donate electrons into the superconducting Fe sheets, and become spectators of the latter.  Unlike the CuO$_2$ sheets in high $T_c$ cuprate superconductors, the Fe sheets do not require O$^{2-}$ ions to become superconducting.  In fact, Ba$_{1-x}$K$_{x}$Fe$_{2}$As$_{2}$ is also a high $T_c$ superconductor with $T_{c}\sim 38$~K \cite{BaK}.    On the other hand, the ground state of the undoped LaOFeAs is magnetically ordered \cite{delaCruz}.  This suggests that the mechanism of superconductivity in LaFeAsO$_{1-x}$F$_{x}$ may be unconventional, and electron-electron correlation effects may be playing a major role, in analogy with high $T_c$ cuprates \cite{Athena}.  More experimental and theoretical efforts are necessary to understand the properties of these fascinating iron-pnictides.

In this {\it Rapid Communication}, we provide a microscopic look at the local electronic properties of LaFeAsO$_{1-x}$F$_{x}$ using NMR (Nuclear Magnetic Resonance).  It is worth recalling that NMR techniques are capable of probing a variety of physical properties of superconductors \cite{MacLaughlin}.  As a start, our present work focuses on the $^{19}$F NMR properties of a polycrystalline sample of LaFeAsO$_{1-x}$F$_{x}$ with the nominal composition $x=0.11$ and $T_{c}\sim28$~K.  See \cite{Athena, Hunte} for comprehensive bulk characterization of the same batch of the sample.  From the measurement of the nuclear spin-lattice relaxation rate, $\frac{1}{T_1}$, we will demonstrate that low frequency spin fluctuations {\it decrease} as the temperature  is lowered toward $T_c$.  The observed behavior is qualitatively similar to that in the {\it pseudo gap} regime of underdoped high temperature cuprate superconductors \cite{Takigawa,Alloul}.  From NMR line broadening induced by vortices in the mixed state below $T_c$, we will also deduce the London penetration depth, $\lambda$.

In Fig.1, we present typical lineshapes of $^{19}$F  NMR observed in an external magnetic field of $B=1.11$~Tesla.  $^{19}$F has nuclear spin $I=\frac{1}{2}$, and is free from complications caused by nuclear quadrupole interactions.  We will take full advantage of the inherently narrow lineshapes to deduce the temperature dependence of $\lambda$ and the intrinsic electron spin susceptibility $\chi$.  Quite generally, one can express the resonance frequency of $^{19}$F NMR as $f = (1+ K)\gamma_{n}B$, where $\gamma_{n}=2\pi \times 40.262$ MHz/Tesla is the nuclear gyromagnetic ratio of the $^{19}$F nuclear spin, and $K$ is the NMR Knight shift,
\begin{equation}
          K = A_{hf}\chi + K_{chem}.
\label{1}
\end{equation}
$A_{hf}$ represents the hyperfine coupling constant between the $^{19}$F nuclear spin and the electron spins, and $K_{chem}$ is a small, temperature independent chemical shift.  From the peak position of the $^{19}$F NMR lineshapes, we deduced the temperature dependence of $K$ as summarized in Fig.2(a).  Superconducting diamagnetic effects and line broadening in the mixed state caused by vortices prevented us from deducing the temperature dependence of $K$ below $T_c$.  $K \lesssim 100$~ppm observed above $T_c$ is comparable to that observed for $^{89}$Y NMR in YBa$_{2}$Cu$_{3}$O$_{x}$ \cite{Alloul}.  The small magnitude of $K$ implies that spin transfer from the Fe sheets to the LaO$_{0.89}$F$_{0.11}$ layers is small, and provides evidence for the highly two dimensional nature of the electronic properties \cite{Singh,Haule}.  Also shown in Fig.2(b) is the temperature dependence of $\frac{1}{T_1T}$, the $^{19}$F nuclear spin lattice relaxation rate $\frac{1}{T_1}$ divided by temperature $T$,
\begin{equation}
          \frac{1}{T_{1}T} \propto |A_{hf}|^{2} \sum_{{\bf k}}{\frac{\chi"({\bf k}, f)}{f}}.
\label{2}
\end{equation}
$\chi"({\bf k}, f)$ represents the imaginary part of the dynamical electron spin susceptibility at the NMR frequency $f$, and the ${\bf k}$ summation is taken over the Brillouin zone.  ($^{19}$F nuclear spin recovery after saturation always showed a single exponential behavior above $T_c$.)    The $^{19}$F site is located directly above/below the Fe sites, and there is no accidental cancellation of antiferromagnetic spin fluctuations for any  ${\bf k}$ modes due to the geometrical form factor (i.e. $|A_{hf}|^{2}$ is independent of ${\bf k}$).  This means that  $\frac{1}{T_1T}$ at the $^{19}$F site probes various ${\bf k}$ modes of low frequency spin fluctuations with equal weight.

A striking feature of Fig.2(b) is that $\frac{1}{T_{1}T}$ monotonically decreases with $T$ from room temperature to $T_c$.  Without relying on any assumptions or model dependent analysis, we conclude that {\it the low frequency spectral weight of spin fluctuations is gradually suppressed with decreasing temperature down to $T_c$}, i.e. the observed suppression establishes the {\it pseudo gap} behavior.  Similar {\it pseudo gap} was observed by NMR and other techniques in the under doped regime of high $T_c$ cuprate superconductors \cite{Takigawa,Alloul,Timusk}, and its possible relation to the superconducting mechanism has been at the focus of intense debate.  Our observation suggests that analogous exotic state of electrons may exist in iron-pnictide above $T_c$.  

Equally interesting is the fact that the Knight shift $K$ in Fig.2(a) shows nearly identical temperature dependence as $\frac{1}{T_{1}T}$ in Fig.2(b).  The inset in Fig.2(b) presents a plot of $\frac{1}{T_{1}T}$ vs. $K$ with temperature as the implicit parameter.  The linear fit suggests that $K_{chem}\sim 30$~ppm and $\frac{1}{T_{1}T} \propto K_{spin} = K - K_{chem} = A_{hf}\chi$.  This simple dynamical scaling behavior implies  $A_{hf} >0$, and is consistent with (but does not necessarily prove) a decreasing $\chi$ with temperature toward $T_c$.  

Next, we turn our attention to the NMR measurements below $T_c$.  As noted above, we are unable to deduce the intrinsic behavior of $K$ below $T_c$, because superconducting vortices induce a distribution of the {\it local} magnetic field $B_{local}$ inside the sample, and the broadening of the NMR lineshape overwhelms the temperature dependence of $K$.  
Nevertheless, one can deduce valuable information about the superconducting state based on this very line broadening, through the measurement of the London penetration depth $\lambda$ as follows \cite{Pincus}.  When our superconducting sample enters the mixed state below $T_c$ in applied magnetic field $B$, quantized magnetic flux $\phi_{o}=\frac{2h}{e}=2.07\times 10^{-7}$ Oe/cm$^2$ forms an Abrikosov lattice.  $B_{local}$ takes the largest value in the vortex cores, and decays with a characteristic length scale set by $\lambda$.  One can express the resulting distribution $\Delta B_{local}$ of the local field as \cite{Brandt}  
\begin{equation}
          \Delta B_{local}  \sim  0.0609 \frac{\phi_{o}}{\lambda^{2}},
\label{3}
\end{equation}
when $B_{c1} \ll B \ll B_{c2}$ is satisfied.  (The lower critical field $B_{c1}$ and upper critical field $B_{c2}$ are estimated to be $\sim 0.005$~Tesla \cite{RenBc1} and $\sim 60$~Tesla \cite{Hunte}, respectively.)  When $\lambda$ begins to decrease from $\lambda = \infty$ below $T_c$, $\Delta B_{local}$ becomes non zero, and induces a temperature dependent broadening of the NMR linewidth, $\Delta f = \gamma_{n} \Delta B_{local}$.  $\Delta f$ is the NMR analogue of the Gaussian relaxation rate, $\sigma$, measured  in $\mu$SR experiments within the time domain \cite{Uemura, Sonier}, and represents the same physical process.  

Experimentally, it is well known that $\Delta B_{local}$ in the superconducting state of polycrystalline samples can be approximated by a Gaussian distribution \cite{Sonier, Heller}.  In fact, we found that $^{19}$F NMR lineshapes fit quite well to Gaussians, as shown in Fig.1.  We deduced $\Delta f$ below $T_c$ from the fit, and used $\Delta f = \gamma_{n} \Delta B_{local}$ and Eq.(3) to convert $\Delta f$ to $\lambda^{-2}$($\propto \Delta f$).  In Fig.3, we summarize the temperature dependence of $\Delta f$ and $\lambda^{-2}$.  One of the most important findings of the present work is that  $\lambda^{-2}$ does not saturate down to the lowest temperature we have reached, $T = 2~K \sim 0.1T_c$.  If LaFeAsO$_{0.89}$F$_{0.11}$ was a conventional BCS s-wave superconductor, $\lambda^{-2}$ would saturate below $T \sim 0.4 T_{c}$, because $\lambda$ would asymptote exponentially to a constant value $\lambda(0)$ below $T \sim 0.4 T_{c}$, where $\lambda(0)$ is the $T=0$ limit.  Instead, $\lambda^{-2}$ continues to grow roughly linearly.

Similar $T$-linear behavior of the Gaussian relaxation rate, $\sigma$, was observed by $\mu$SR for clean polycrystalline samples of high $T_c$ superconductors, and it is known to be an indication of the presence of line nodes in the superconducting energy gap $\Delta ({\bf k})$ \cite{Sonier}.  The underlying physics of such T-linear behavior is easily understandable from the following considerations.  When there are {\it line nodes} in $\Delta ({\bf k})$ at certain wave vector ${\bf k}$, the density of states of superconducting carriers averaged over ${\bf k}$ increases linearly with energy, $D_{s}(E-E_{F}) \sim |E - E_{F}|$.  For this reason, the superconducting carrier density $n_{s}$ decreases linearly above $T=0$.  In the London theory, $n_{s}$ and $\lambda^{-2}$ are proportional to each other,  
\begin{equation}
          n_{s}=\frac{m^{*}c^{2}}{4\pi e^{2} \lambda^{2}},
\label{4}
\end{equation}
where $m^{*}$ is the effective mass of the carrier.  Eq.(4) indicates that $\lambda^{-2}$ shows a T-linear behavior near $T=0$ if $\Delta ({\bf k})$ has line nodes.  

Close inspection of our $\lambda^{-2}$ data in Fig.3 reveals a small, negative curvature.   It is known that disorder tends to smear out the line nodes, and $\lambda^{-2}$ begins to deviate from T-linear behavior, crossing over to $T^2$ behavior when the sample is strongly disordered \cite{Hirschfeld}.  We find similarity in the temperature dependence of our $\lambda^{-2}$ data and theoretical calculations  based on a d-wave model with mild disorder, as shown in Fig.3.  We emphasize, however, that this qualitative agreement does not necessarily mean that the superconducting energy gap in LaFeAsO$_{0.89}$F$_{0.11}$ has d-wave symmetry.  Our results for $\lambda^{-2}$ alone cannot discern \cite{Hirschfeld2} d-wave symmetry from extended s-wave \cite{Mazin} or p-wave \cite{Lee}.

By smoothly extrapolating the temperature dependence of $\lambda^{-2}$ to $T=0$, we estimate the $T=0$ limit as $\lambda(0)^{-2}\sim 7$ $\mu$m$^{-2}$, or equivalently,  $\lambda(0) \sim 378$ nm.  Our estimation of the London penetration depth $\lambda(0)$ is averaged over all possible orientations.  As noted above, however, LaFeAsO$_{1-x}$F$_{x}$ should be considered a highly anisotropic system, hence we expect an anisotropic penetration depth, $\lambda_{ab} \ll \lambda_{c}$.  Theoretically, the in-plane penetration depth $\lambda_{ab}$ and the powder averaged penetration depth $\lambda$ satisfy a simple relation,  $\lambda_{ab} \sim \lambda / 1.23$ \cite{Braford}, hence we arrive at $\lambda_{ab}(0) \sim 308$~nm.  This is comparable to $\lambda_{ab}(0)=250$~nm observed for the high $T_c$ cuprate La$_{1.85}$Sr$_{0.15}$CuO$_{4}$ with comparable $T_{c}=38$~K \cite{Sonier}.

The observed value of $\lambda(0)$ also provides additional insight into the electronic properties of the LaFeAsO$_{1-x}$F$_{x}$ system.  First, combined with the superconducting correlation length $\xi(0)\sim 4$~nm deduced from the zero temperature limit of the upper critical field $B_{c2}(0)$ \cite{Hunte}, we arrive at the Ginzburg-Landau parameter $\kappa = \lambda(0)/\xi(0) \sim 77$.  Second, within the London relation, Eq.(3), and ignoring possible mass enhancement for $m^{*}$, we  estimate the density of superconducting carriers as $n_{s} \sim 0.3 \times 10^{21}$ cm$^{-3}$.  This is comparable to the carrier density estimated by Hall voltage measurements in the normal state above $T_c$, $n_{H}\sim 1.0 \times 10^{21}$ cm$^{-3}$ \cite{Athena}.  As already noted in Ref.\cite{Athena}, the LaFeAsO$_{1-x}$F$_{x}$ system is indeed a low carrier density superconductor.   

Given the extraordinarily fast progress of research into iron-pnictide systems, it would be useful to compare our findings with those by other experimental techniques, and put our results into perspective.  During the course of the present study, measurements of the field dependence of the specific heat \cite{Mu}, tunneling conductance \cite{Shan}, lower critical field $B_{c1}$ \cite{RenBc1}, and $\mu$SR \cite{Luetkens} were independently reported, all pointing toward the presence of line nodes in the energy gap, in agreement with our conclusion.  In particular, Ren et al. reported a linear temperature dependence of $B_{c1}$ below $T\sim0.5T_c$ \cite{RenBc1}.  Within the Ginzburg-Landau theory, $\lambda^{-2} \propto B_{c1}$, hence the $B_{c1}$ and NMR measurements are fully consistent.  The penetration depth $\lambda_{ab}(0) = 254$~nm as measured by by $\mu$SR \cite{Luetkens} is also in good agreement with our finding.  Very recently, $^{139}$La NMR \cite{Nakai}, $^{75}$As NMR \cite{Nakai, Grafe}, and photoemission experiments \cite{photo} confirmed the pseudo gap behavior above $T_c$.  Moreover, $\frac{1}{T_1T}$ data at $^{75}$As and $^{139}$La sites show an absence of the Hebel-Slichter coherence peak just below $T_c$, and exhibit a power-law temperature dependence well below $T_c$ \cite{Nakai,Grafe}.  These results are consistent with the presence of line nodes in $\Delta ({\bf k})$. 

To summarize, we have reported a microscopic investigation of the local electronic properties of the LaFeAsO$_{0.89}$F$_{0.11}$ superconductor based on $^{19}$F NMR techniques.  Our observations indicate that LaFeAsO$_{0.89}$F$_{0.11}$ is a highly unconventional system. 

T.I. thanks G. M. Luke and A. J. Berlinsky for helpful discussions.  The work at McMaster was supported by NSERC and CIFAR.  Research sponsored by the Division of Materials Science and Engineering, Office of Basic Sciences, Oak Ridge National Laboratory is managed by UT-Battelle, LLC, for the U.S. Department of Energy under contract No. DE-AC-05-00OR22725.\\


Fig.1. Left : Crystal structure of LaFeAsO$_{0.89}$F$_{0.11}$.   Right : Representative $^{19}$F NMR lineshapes for a randomly oriented polycrystalline sample.  Curves represent a Gaussian fit.  \\

Fig.2. (a) Temperature dependence of the NMR Knight shift $K$.  (b) Spin-lattice relaxation rate divided by $T$, $\frac{1}{T_{1}T}$.   Inset to (b) : $1/T_{1}T$ vs $K$ with temperature as the implicit parameter.   The linear fit suggests that $1/T_{1}T$ and $K$ exhibit the same temperature dependence, and $K_{chem}=30$~ppm.  The solid curves in (a) and (b) represent empirical parabolic fits with the same curvature and $K_{chem}=30$~ppm.  \\

Fig.3 Temperature dependence of the broadening of the Gaussian linewidth $\Delta f$ below $T_c$.  The right axis shows the conversion to $\lambda^{-2}$ (see main text).  Curves represent a theoretical fit to an energy gap with d-wave symmetry,  without disorder (solid curve) and with a mild disorder $\Gamma/T_{c}=0.01$ (dashed curve) and $\Gamma/T_{c}=0.05$ (dotted curve), where $\Gamma$ is the scattering rate \cite{Hirschfeld}. \\

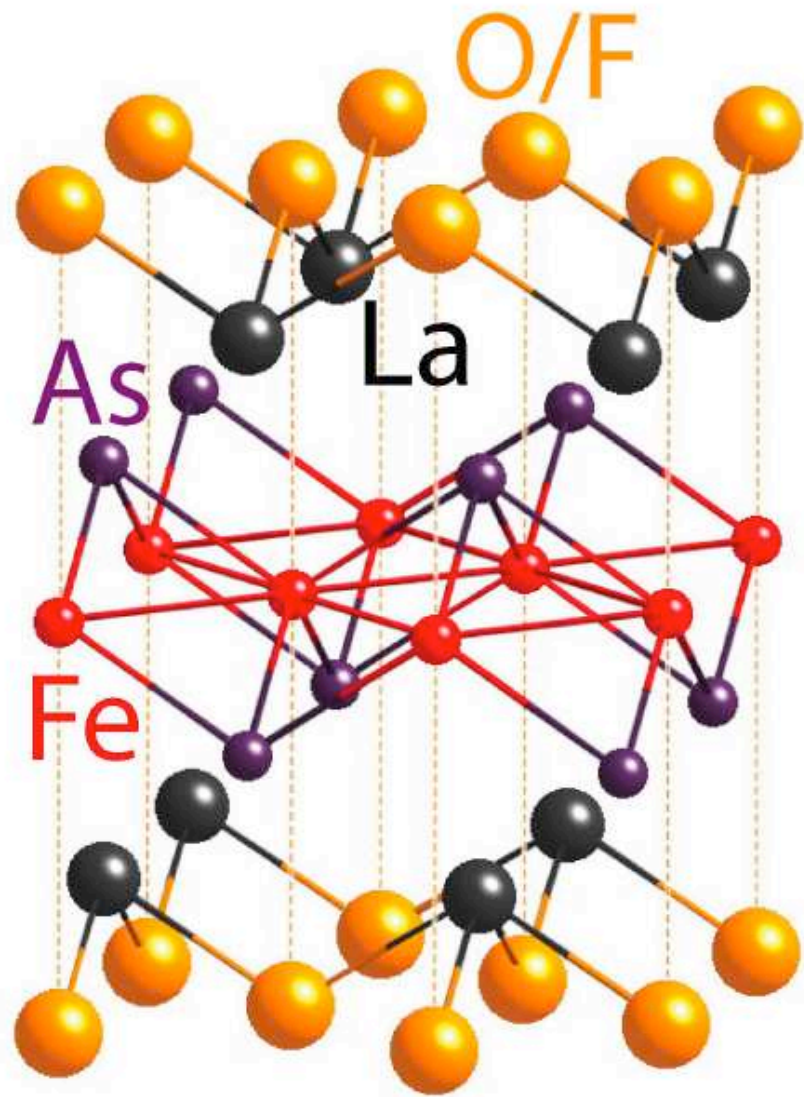
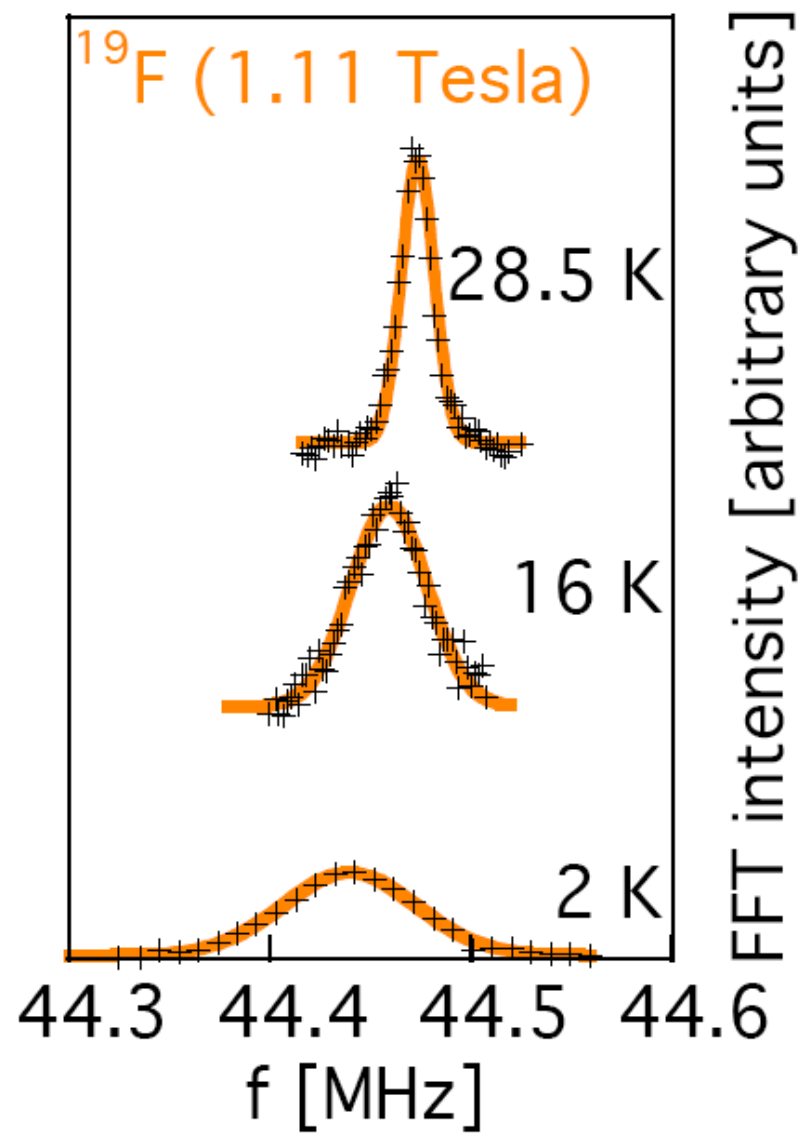
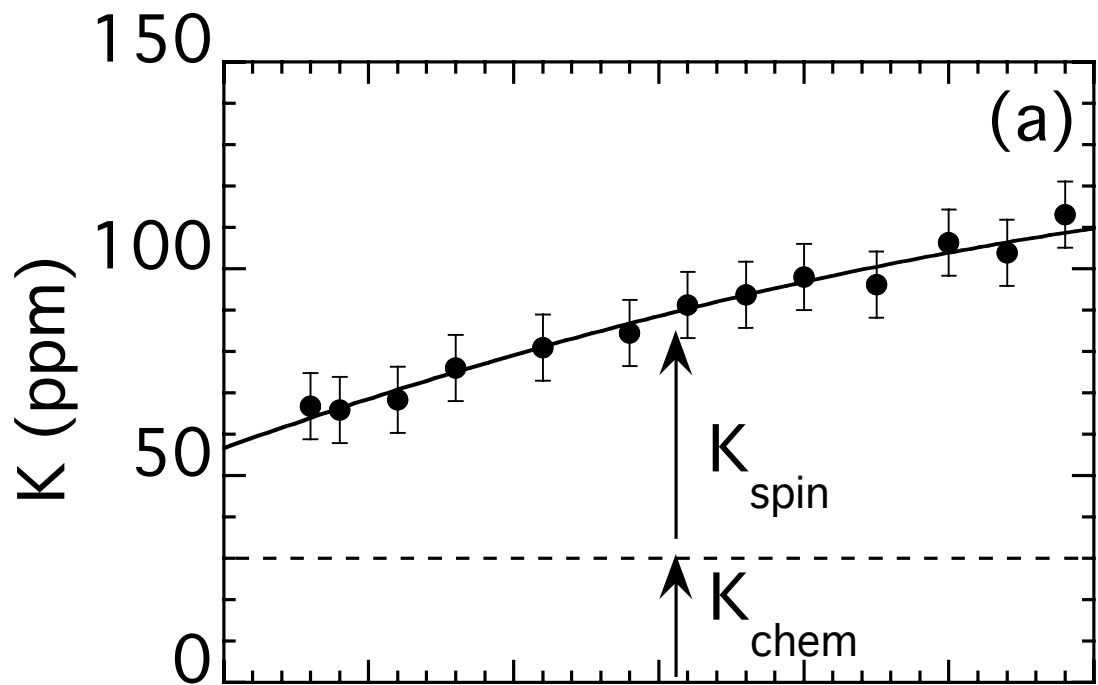
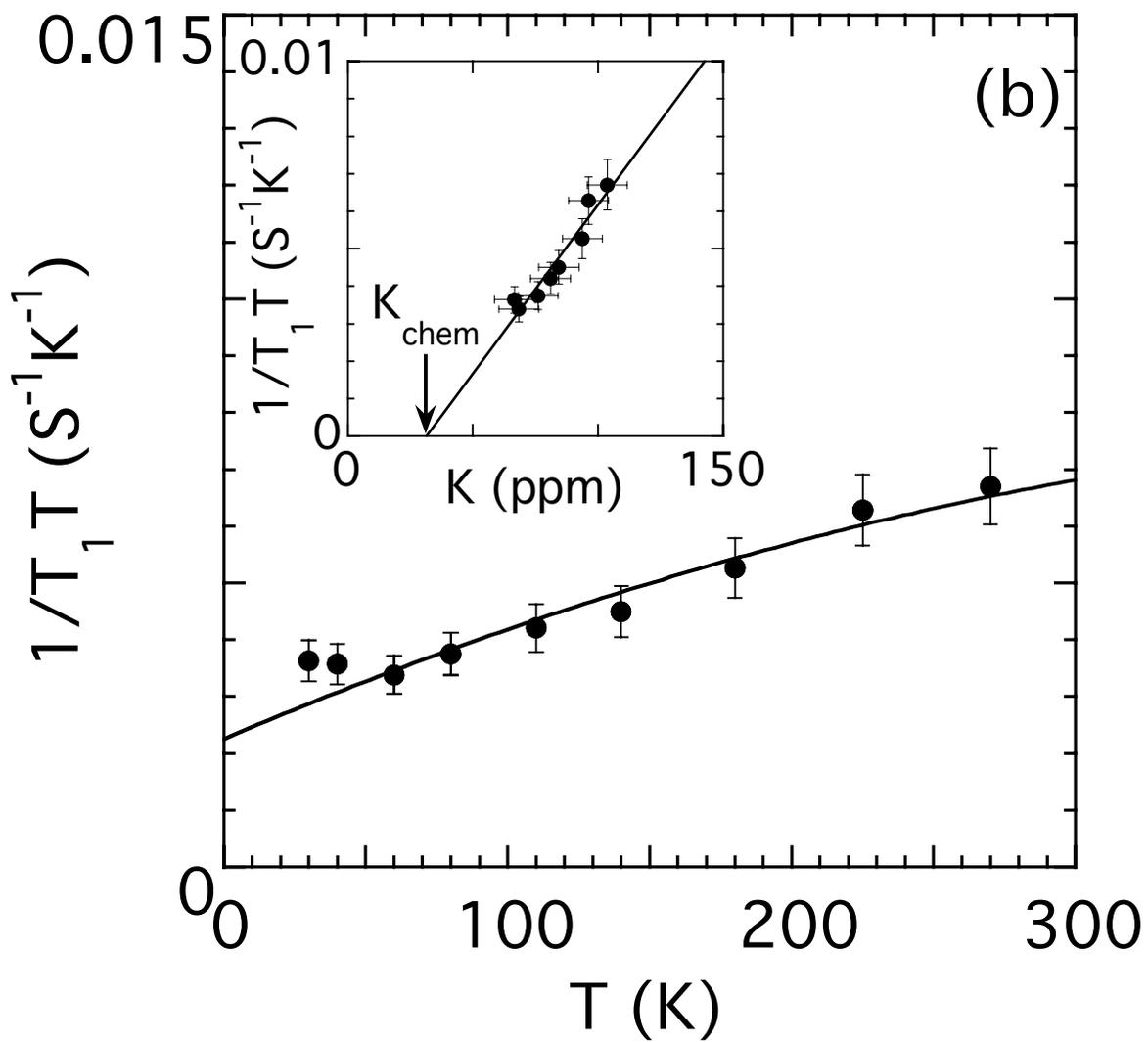
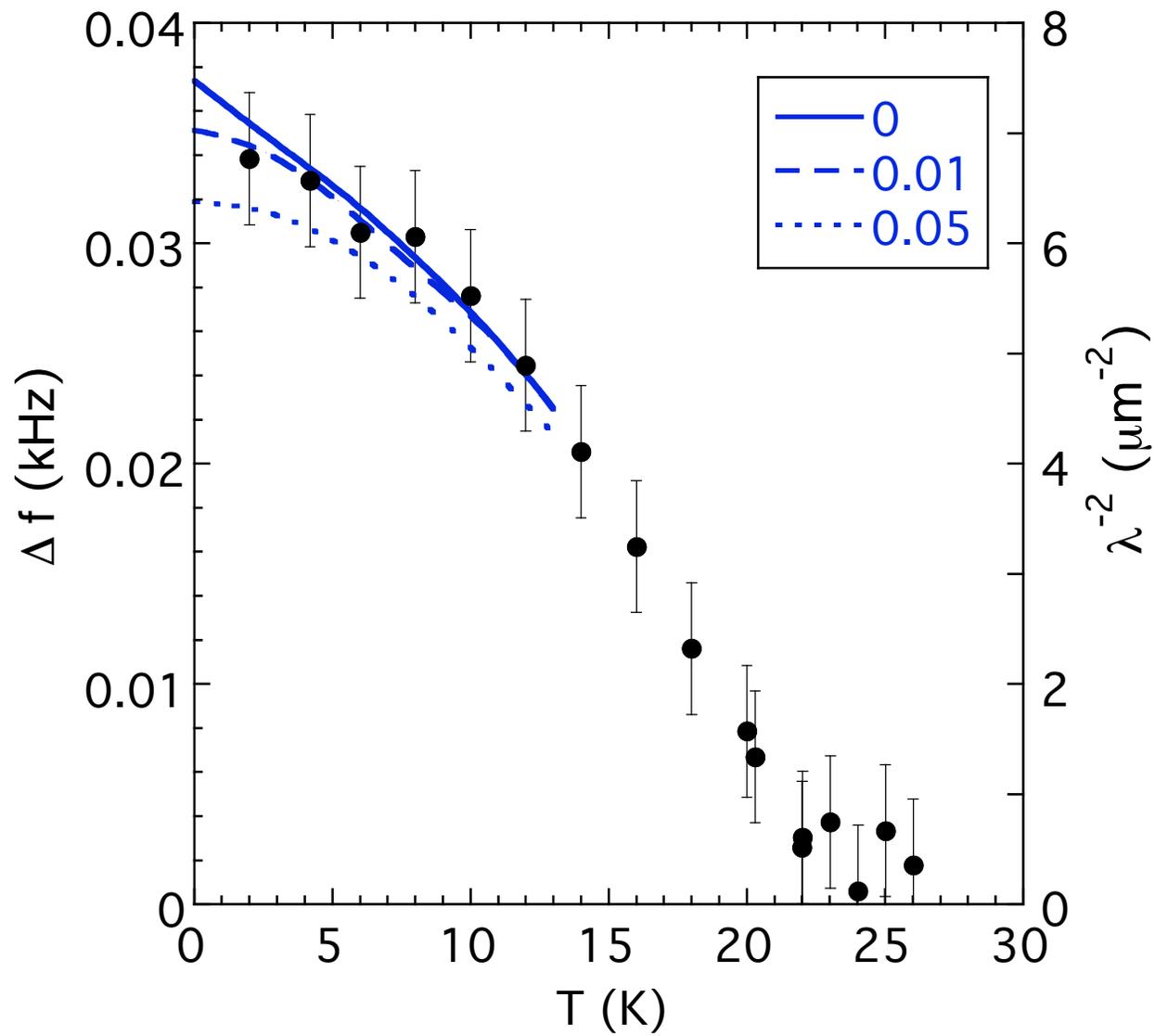
\end{document}